\documentclass[10pt,twoside,twocolumn,english,aps,prl,superscriptaddress,notitlepage]{revtex4-2}
\usepackage{helvet}

\usepackage[latin9]{inputenc}
\usepackage[letterpaper]{geometry}
\usepackage{color}
\usepackage{babel}
\usepackage{verbatim}
\usepackage{textcomp}
\usepackage{amsbsy}
\usepackage{amsmath}
\usepackage{amstext}
\usepackage{graphicx}
\usepackage{esint}
\usepackage{natbib}
\usepackage{multirow}
\usepackage[unicode=true,pdfusetitle,
 bookmarks=false,
 breaklinks=true,pdfborder={0 0 0},pdfborderstyle={},backref=false,colorlinks=true]
 {hyperref}
\hypersetup{
 colorlinks=true,citecolor=blue,linkcolor=black,urlcolor=blue,pagebackref=true}

\ProvideTextCommand{\~}{LGR}[1]{\char126#1}
\newcommand{\lyxmathsym}[1]{\ifmmode\begingroup\def\b@ld{bold}
  \text{\ifx\math@version\b@ld\bfseries\fi#1}\endgroup\else#1\fi}

\usepackage{amsfonts,anyfontsize,xcolor,graphicx}
\usepackage[calcwidth,explicit]{titlesec}
\usepackage[normalem]{ulem}

\titleformat{\section}{\bfseries\large\sffamily\scshape\filcenter}{\thesection.}{0.2em}{#1}
\titlespacing{\section}{0pt}{0.6ex}{0.2ex}
\titleformat{\paragraph}[runin]{\normalfont\normalsize\bfseries}{}{0pt}{\theparagraph}
\titlespacing*{\paragraph}{0em}{0ex}{0.3em}[]

\makeatletter\renewcommand\frontmatter@abstractwidth{\dimexpr\textwidth-2cm\relax}\makeatother

\addto\captionsenglish{\renewcommand{\figurename}{FIG.}}

\renewcommand\thesection{\Alph{section}}

\makeatletter\@addtoreset{paragraph}{section}\makeatother
\makeatletter\def\p@paragraph{}\makeatother
\makeatletter\AtBeginDocument{\let\@elt\relax}\makeatother

\AtBeginDocument{
\renewcommand{\ref}[1]{\autoref{#1}}

}

\makeatother

\begin{document}
\title{Skyrmion Generation from Irreversible Fission of Stripes in Chiral
Multilayer Films\smallskip{}
}
\author{Anthony K.C. Tan}
\affiliation{Data Storage Institute, Agency for Science, Technology \& Research, 138634 Singapore}
\author{James Lourembam}
\thanks{These authors contributed equally to this work.}
\affiliation{Institute of Materials Research \& Engineering, Agency for Science,
Technology \& Research, 138634 Singapore}
\affiliation{Data Storage Institute, Agency for Science, Technology \& Research, 138634 Singapore}
\author{Xiaoye Chen}
\thanks{These authors contributed equally to this work.}
\affiliation{Institute of Materials Research \& Engineering, Agency for Science,
Technology \& Research, 138634 Singapore}
\affiliation{Data Storage Institute, Agency for Science, Technology \& Research, 138634 Singapore}
\author{Pin Ho}
\affiliation{Institute of Materials Research \& Engineering, Agency for Science,
Technology \& Research, 138634 Singapore}
\affiliation{Data Storage Institute, Agency for Science, Technology \& Research, 138634 Singapore}
\author{Hang Khume Tan}
\affiliation{Institute of Materials Research \& Engineering, Agency for Science,
Technology \& Research, 138634 Singapore}
\affiliation{Data Storage Institute, Agency for Science, Technology \& Research, 138634 Singapore}
\author{Anjan Soumyanarayanan}
\email{anjan@imre.a-star.edu.sg}

\affiliation{Institute of Materials Research \& Engineering, Agency for Science,
Technology \& Research, 138634 Singapore}
\affiliation{Data Storage Institute, Agency for Science, Technology \& Research, 138634 Singapore}
\affiliation{Physics Department, National University of Singapore, 117551
Singapore }
\begin{abstract}
\noindent Competing interactions produce finite-sized textures in myriad condensed matter systems -- typically forming elongated ``stripe'' or round ``bubble'' domains. Transitions between stripe and bubble phases, driven by field or temperature, are expected to be reversible in nature. Here we report on the distinct character of the analogous transition for nanoscale spin textures in chiral Co/Pt-based multilayer films - known to host N\'{e}el skyrmions - using microscopy, magnetometry, and micromagnetic simulations. Upon increasing field, individual stripes fission into multiple skyrmions, and this transition exhibits a macroscopic signature of irreversibility. Crucially, upon field reversal, the skyrmions do not fuse back into stripes -- with many skyrmions retaining their morphology down to zero field. Both the macroscopic irreversibility and the microscopic zero field skyrmion density are governed by the thermodynamic material parameter determining chiral domain stability. These results establish the thermodynamic and microscopic framework underlying ambient skyrmion generation and stability in chiral multilayer films and provide immediate directions for their functionalization in devices.
\end{abstract}
\maketitle

\noindent \section{Introduction}

\paragraph{Sk Intro}
Magnetic skyrmions are nanoscale, topologically wound spin structures recently stabilized at room temperature (RT) in industry-friendly multilayer thin films \citep{MoreauLuchaire2016,Boulle2016,Woo2016,Soumyanarayanan2017,Fert2017}. Their particle-like properties have spawned synergistic efforts to investigate their nucleation and dynamics, and to harness them for practical spintronic devices \citep{Romming2013,Jiang2015,Woo2016,Buttner2017,Soumyanarayanan2016}. Imperative to the utility of multilayer skyrmions is a comprehensive understanding of their creation mechanism and stability at zero magnetic field (ZF) with respect to proximate magnetic states \citep{Wiesendanger2016,Fert2017}.

\paragraph{Sk Thermodynamics}
Skyrmions may form due to the competition between symmetric exchange and anti-symmetric Dzyaloshinskii-Moriya interaction (DMI) -- which prefer uniform and spiral spin arrangements respectively \citep{Bogdanov2001,Nagaosa2013,Wiesendanger2016}. While their much-touted topology provides an additional energy barrier, it is far from the only factor determining stability \citep{CortesOrtuno2017,Buttner2018}. Within the Ginzburg-Landau picture for modulated phases \citep{Seul1995}, the thermodynamic stability of skyrmions in chiral magnets is characterized by the material parameter $\kappa$ \citep{Bogdanov2001,Rohart2013,Leonov2016}
\begin{equation}
\kappa=\frac{\pi{}D}{4\sqrt{AK_{{\rm eff}}}}\label{eq:Kappa_form}
\end{equation}
Here, $D$ is the DMI strength, $A$ is the exchange stiffness and $K_{{\rm eff}}$ is the effective anisotropy. For $\kappa>1$ (stable limit), skyrmions form a lattice at finite fields, which at reduced fields should elongate or ``strip-out'' to give a labyrinthine stripe (LS) state at remanence \citep{Kiselev2011,Leonov2016}. As is the case for conventional modulated phases \citep{Seul1995,Saratz2010}, the strip-out transition for chiral spin textures is expected to be reversible with magnetic field. However, elucidating the skyrmion-stripe transition and its relevance to ZF skyrmion stability is challenging, as it requires direct correspondence between microscopic and thermodynamic regimes.

\paragraph{FORC Technique}
First order reversal curve (FORC) techniques form an invaluable bridge linking the microscopic evolution of magnetization ($M$) with thermodynamic signatures. A FORC is defined as a field segment of a minor hysteresis loop, $M(H,H_{{\rm r}})$ (\ref{fig:FORC-Motivation}a: gray lines), and is characterized by a reversal field, $\mu_{0}H_{{\rm r}}$ (\ref{fig:FORC-Motivation}a: circles) \citep{Pike1999,Roberts2000}. A set of FORCs acquired at regular $H_{{\rm r}}$ intervals enable the determination of the irreversibility, $\rho (H,H_{\rm r})$ (see \ref{eq:Rho_Form}). A finite value of $\rho$ is associated with irreversible magnetic switching \citep{Pike1999,Davies2004,Davies2008,Kirby2010,Bonanni2010}. FORC magnetometry has previously shed light on myriad magnetic phenomena -- e.g. mineral phases \citep{Roberts2000}, nanomagnetic interactions \citep{Pike1999,Gilbert2014}, and magnetic domains \citep{Davies2004}. Notably, FORC, complemented by magnetic microscopy, has been instrumental in elucidating domain nucleation and annihilation mechanisms in Co-based multilayer films \citep{Davies2004,Davies2008,Kirby2010}. Such transitions involve the irreversible creation or rupture of domain walls, and therefore manifest as distinct peaks in $\rho$ \citep{Davies2004,Davies2008,Kirby2010}. However, the sensitivity of FORC to transitions between distinct magnetic textures remains to be established.

\paragraph{Results Summary}
Here we report on chiral spin textures at remanence, and their evolution with reversal field in Co/Pt-based multilayer films with varying $\kappa$ -- using FORC magnetometry, magnetic force microscopy (MFM) and micromagnetic simulations. For $\kappa>1$ samples, FORC shows -- in addition to conventional near-saturation features -- a hitherto unpredicted peak at intermediate fields. Imaging and simulations reveal that this peak originates from the fission of individual stripes into multiple skyrmions. Crucially, upon reversing the field these skyrmions do not fuse back into stripes, which results in substantial enhancement of the ZF skyrmion density for $\kappa>1$ samples. Finally, both the unconventional FORC feature and ZF skyrmions diminish as $\kappa$ is reduced, and disappear for $\kappa$ below unity. Our work provides a comprehensive platform to engineer skyrmions stable at ZF, and a mechanistic link between microscopic and thermodynamic characteristics of chiral spin textures. 
\noindent \section{Methods\label{sec:Methods}}

\paragraph{Film Deposition}
Multilayer thin films of Ta(40)/Pt(50)/{[}Ir(10)/Fe($x$)/ Co($y$)/Pt(10){]}$_{14}$/Pt(20) (thickness in angstroms in parentheses) were deposited on 100 mm SiO$_{2}$/Si wafers by DC magnetron sputtering at RT using a Chiron\texttrademark{} UHV system (base pressure $<5\times10^{-8}$~torr) from Bestec GmbH. Ir/Fe($x$)/Co($y$)/Pt multilayers has been established as a platform for realizing N\'{e}el-textured RT skyrmions with smoothly tunable sizes ($\sim$30-100~nm) and densities ($\sim$5-60~$\mu$m$^{-2}$) \citep{Soumyanarayanan2017,Yagil2018,Ho2019}. The five samples studied here are described henceforth in terms of their Fe($x$)/Co($y$) composition: Fe(0)/Co(10), Fe(2)/Co(8), Fe(3)/Co(7), Fe(4)/Co(6), and Fe(5)/Co(5). Together, they allow us to modulate the DMI, $D$ over 0.9--2.1~mJ/m$^{2}$ and anisotropy, $K_{{\rm eff}}$ over 0.01--0.60~MJ/m$^{3}$, while the saturation magnetization, $M_\mathrm{S}$ remains constant to within 20\% \citep{SuM,Johnson_1996}. As a result, $\kappa$ varies by over an order of magnitude (0.3 - 4.1) across the samples\citep{Soumyanarayanan2017}. To characterize the samples using microscopy and FORC magnetometry techniques, magnetic fields were applied out-of-plane, and are herein referenced to the saturation field ($\mu_{0}H_{{\rm s}}$).

\paragraph{Magnetic Microscopy}
MFM measurements were performed with a Veeco Dimension\texttrademark{} 3100 microscope. Similar to our previous studies \citep{Ho2019}, ultra-low moment SSS-MFMR\texttrademark{} tips by NANOSENSORS\texttrademark{} (remanent magnetization $\sim80$~emu/cm$^{3}$, diameter $\sim30$~nm) were used with lift heights of 15-30~nm to obtain high spatial resolution MFM images with minimal stray field perturbation. The FORC sequence was replicated for the MFM imaging: each sample was imaged after \emph{ex situ} negative OP saturation ($H<-H_{{\rm s}}$) followed by (i) the application of \emph{in situ} positive OP fields ($H\equiv H_{{\rm r}}$), and (ii) at zero field after subjecting to $H_{{\rm r}}$ in (i). Skyrmions were identified within MFM images using established image processing techniques \citep{SuM,Otsu1979}. 

\paragraph{FORC Magnetometry}
A Micromag\texttrademark{} Alternating Gradient Magnetometer (AGM) was used to characterize diced 4 \texttimes 4~mm$^{2}$ samples at RT. The major hysteresis loop was first measured for each sample to determine appropriate parameters for FORC measurements. For FORC loops, the samples were first saturated at $-H_{0}=-500$~mT (over $1.5\,H_{{\rm s}}$). Subsequently, the field was increased from $-H_{0}$ to the reversal field $H_{{\rm r}}$, and a FORC was traced by measuring $M(H)$ as the field was brought back to $-H_{0}$. A set of FORCs were acquired with this recipe at steps of 2-3~mT ($H$ and $H_{{\rm r}}$) with 0.1~s of averaging per field point \citep{SuM}. The datasets acquired for each sample comprise over 300 FORCs. The FORC dataset was used to determine the irreversibility, $\rho (H,H_{\rm r})$, defined as \citep{Pike1999,Roberts2000,Davies2004,Davies2008,Kirby2010,Bonanni2010,Gilbert2014}
\begin{equation}
\rho(H,H_{{\rm r}})=-\frac{1}{2}\,\frac{\partial^{2}M(H,H_{{\rm r}})}{\partial H\,\partial H_{{\rm r}}}\quad.\label{eq:Rho_Form}
\end{equation}

Here, $\rho (H,H_{\rm r})$ was calculated based on a well-established method of fitting the magnetization to a second order polynomial surface \citep{Pike1999}
\begin{equation}
M(H,H_{{\rm r}})=a_{1}+a_{2}H_{{\rm r}}+a_{3}H+a_{4}H_{{\rm r}}^{2}+a_{5}H^{2}+a_{6}H_{{\rm r}}H\label{eq:FORC_MHFit}
\end{equation}
In this case, the magnitude of $\rho(H,H_{{\rm r}})$ (see \ref{eq:Rho_Form}) is given by $-a_{6}$, which is plotted in \ref{fig:FORC-Expts}, \ref{fig:FORC-Sims}, \ref{fig:Irrev-SampleDep}. The number of magnetization points used for each fit is $(2\cdot{\rm SF}+1)^{2}$, where ${\rm SF}$, the smoothing factor, is 3 for this work. The FORC analysis techniques used here have been previously employed in several studies \citep{Davies2004,Bonanni2010,Gilbert2014}.

\paragraph{Simulations}
Micromagnetic simulations were performed on the full 14-repeat stack structure using mumax\textthreesuperior{}\citep{Vansteenkiste2014}, which natively accounts for DMI, magnetostatic interactions, and entropic effects \citep{SuM,Brown_1993,Lyberatos_1993}. The simulated area was 1.5~$\mu$m in size with a 4~nm mesh, and the effective medium approximation with one layer per repetition is used to optimize memory and computing requirements \citep{Woo2016}. The magnetic parameters ($M_{{\rm s}}$, $K_{{\rm eff}}$, $A$, $D$) used for the simulations are obtained using protocols consistent with literature \citep{MoreauLuchaire2016,Woo2016,Soumyanarayanan2017,Ho2019}, and are detailed in \citep{SuM}. Major hysteresis loops as well as FORC loops with varying reversal fields (shown in \citep{SuM}) are simulated using methods similar to those described in \citep{Vansteenkiste2014}. The irreversibility $\rho(H,H_{{\rm r}})$ is calculated using the polynomial fit method described above.

\noindent \section{Remnant Textures \& Field Reversal\label{sec:FORC-Expts}}

\begin{figure}
\noindent \begin{centering}
\includegraphics[width=3.3in]{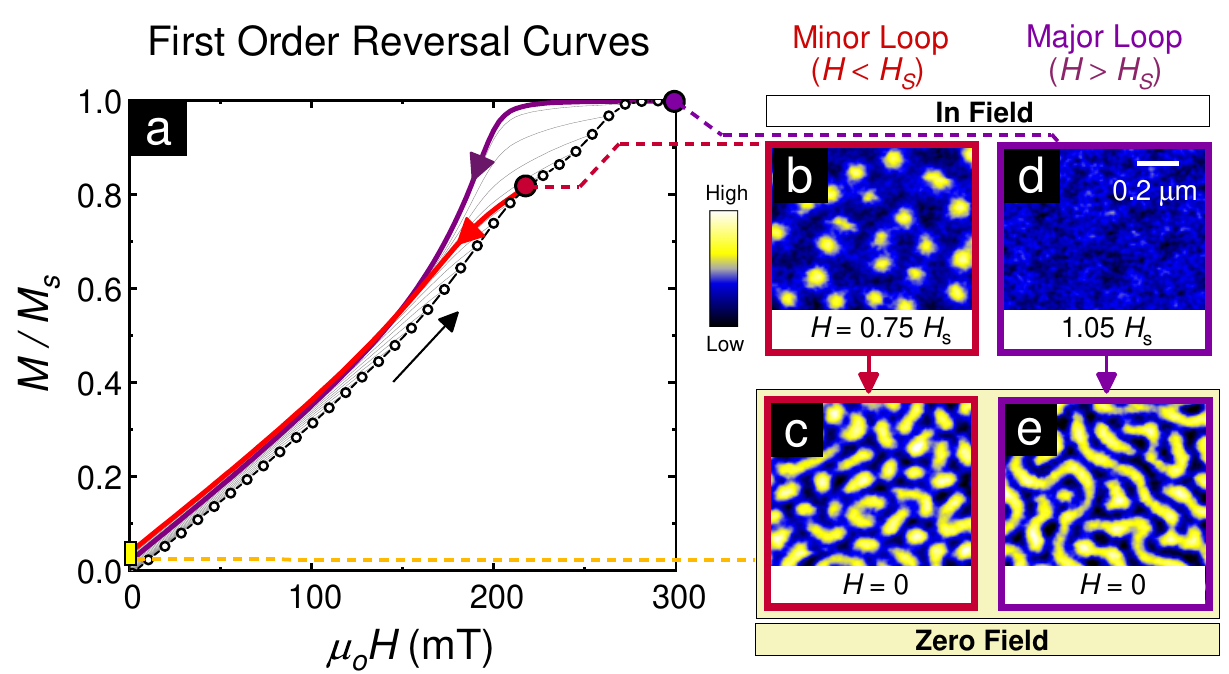}
\par\end{centering}
\noindent \caption[Hysteretic ZF Textures]{\textbf{Hysteretic Nature of Zero Field (ZF) Textures in Fe(4)/Co(6). (a)} A set of first-order reversal curves (FORCs, grey lines) of sample Fe(4)/Co(6) -- acquired by sweeping field from $-H_{\rm s}$ to the reversal field $H_{\rm r}$ (white circles), and back to $-H_{\rm s}$. Arrows indicate FORC field sweep protocol up to $H_{\rm r}$ (black), and the reversal path for representative major (purple, $H_{\rm r} > H_{\rm s}$) and minor (red, $H_{\rm r} < H_{\rm s}$) loops. \textbf{(b-e)} Also shown are the MFM images for two selected FORCs (red and purple dots in (a)), showing in (b) and (d) the in-field (IF) and in (c) and (e) the zero-field (ZF) configurations. The two cases shown represent: (b,c) a minor loop ($H_{{\rm r}}=220$ ~mT $<H_{\rm s}$), corresponding to an IF skyrmion phase; and (d,e) a major loop ($H_{{\rm r}}=310$ ~mT $>H_{{\rm s}}$), i.e. a uniformly magnetized IF phase. \label{fig:FORC-Motivation}}
\end{figure}

\paragraph{Hysteretic Morphology}
The key empirical motivation for this work relates to the hysteretic morphology of magnetic textures for Fe(4)/Co(6) ($D\simeq1.9$~mJ/m$^{2}$; $\kappa\simeq1.6$), which hosts thermodynamically stable skyrmions \citep{Soumyanarayanan2017,Ho2019}. MFM imaging of the conventional ZF state -- obtained along the major loop (\ref{fig:FORC-Motivation}(a): purple), i.e. by reversing from uniform magnetization (\ref{fig:FORC-Motivation}(d)) -- shows a labyrinthine stripe phase (\ref{fig:FORC-Motivation}(e)) -- consistent with results on 20x multilayers \citep{Soumyanarayanan2017,Ho2019,Raju2019}. In contrast, the remnant state is drastically altered upon following a minor loop, e.g. by reversing from $H_{{\rm r}}\sim0.8H_{{\rm s}}$ (\ref{fig:FORC-Motivation}(a): red), wherein the in-field (IF: $H=H_{{\rm r}}$) state consists of skyrmions (\ref{fig:FORC-Motivation}(b)). In this case, the ZF state instead comprises skyrmions and short magnetic stripes (\ref{fig:FORC-Motivation}(c)). This minor loop field segment, when extended from $H_{{\rm r}}$ to $-H_{{\rm s}}$, constitutes a FORC \citep{Pike1999,SuM}. This correspondence motivates the use of FORC magnetometry to understand the observed hysteretic stability of skyrmions at ZF \citep{Desautels2019,Herve2018,Gross2018,Zeissler2017}.

\noindent
\begin{figure}
\noindent \begin{centering}
\includegraphics[width=3.3in]{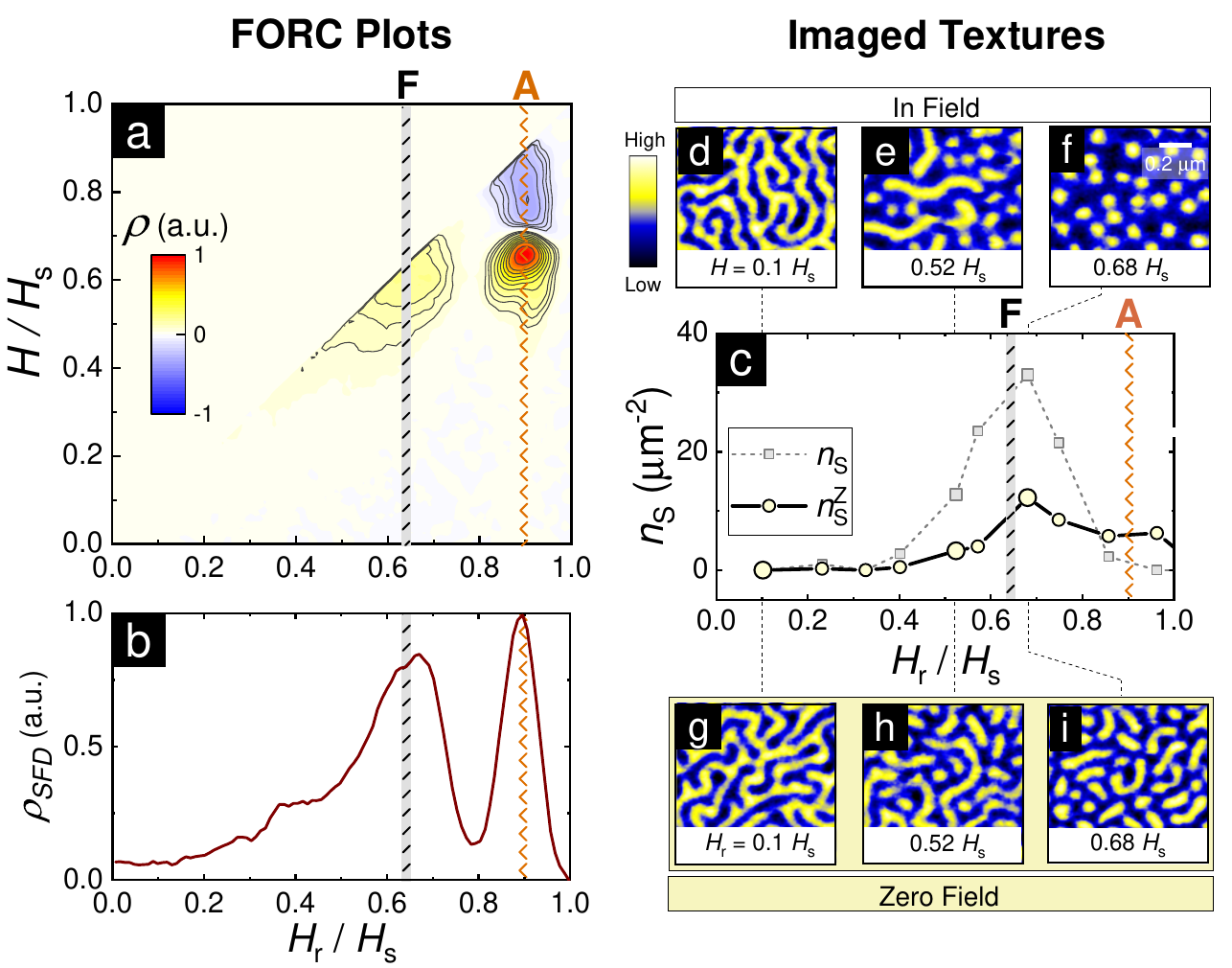}
\par\end{centering}
\noindent \caption[Experimental FORC \& ZF Textures]{\textbf{Experimental FORC Irreversibility and ZF Textures for Fe(4)/Co(6) stacks. (a)} Color plot of the measured FORC irreversibility, $\rho(H,H_{{\rm r}})$ (defined in \ref{eq:Rho_Form}) and  \textbf{(b)} the switching field distribution (SFD, defined in \ref{eq:SFD_Form}) obtained from set of FORCs on Fe(4)/Co(6) (full FORCs in \citep{SuM}). The two prominent irreversible processes ($\rho\protect\neq0$), marked by black and orange lines, are denoted as ${\mathcal{F}}$ and ${\mathcal{A}}$ respectively (details in text). $\mathcal{F}$ and $\mathcal{A}$ indicators are also shown within imaging results for direct comparison with FORC.\textbf{ (c) }Corresponding skyrmion density plots from MFM imaging of IF ($n_{{\rm S}}$) and ZF ($n_{{\rm S}}^{Z}$) morphology with varying $H$ and $H_{{\rm r}}$ respectively.\textbf{ (d-i)} Representative MFM images of IF (d-f) and ZF (g-i) morphology for selected $(H,H_{{\rm r}})$ pairs (enlarged data markers in (c)). \label{fig:FORC-Expts}}
\end{figure}

\paragraph{New $\rho$ Feature in FORC}
\ref{fig:FORC-Expts}(a) shows the FORC irreversibility distribution, $\rho(H,H_{{\rm r}})$, for \textbf{Fe(4)/Co(6)} plotted as a function of $H$ and $H_{{\rm r}}$. While $\rho(H,H_{{\rm r}})$ is predominantly zero in such plots \citep{SuM}, features of interest are clusters of non-zero $\rho$ values -- indicative of distinct irreversible processes \citep{Davies2004,Davies2008,Bonanni2010,Kirby2010}. An inspection of \ref{fig:FORC-Expts}(a) reveals 2 distinct features -- at $\sim0.9\,H_{{\rm s}}$ and $\sim0.6\,H_{{\rm s}}$ respectively. First, the peak-valley feature at $H_{{\rm r}}\sim0.9\,H_{{\rm s}}$ is well-studied in conventional Co-based multilayer systems \citep{Davies2004,Davies2008,Kirby2010}. Labeled as $\mathcal{A}$ (\ref{fig:FORC-Expts}(a): orange line), this pair feature is known to arise from the annihilation of domains near saturation, which manifests as the divergence of neighbouring FORCs, followed by their bunching (see \ref{fig:FORC-Motivation}(a))\citep{Davies2004,Davies2008,Kirby2010}. However, more interesting is the broad peak centered at $\sim0.6\,H_{{\rm s}}$ adjacent to the $H=H_{{\rm r}}$ edge. Suggestively labeled as $\mathcal{F}$ (\ref{fig:FORC-Expts}(a): black line), this unexpected feature will subsequently be established as a thermodynamic signature of the stripe-skyrmion transition. Meanwhile, in line with previous FORC work, we can project the FORC distribution onto the $H_{{\rm r}}/H_{{\rm s}}$ axis to plot the switching field distribution (SFD), defined as \citep{Davies2004,Bonanni2010,Kirby2010}:

\noindent
\begin{equation}
\rho_{{\rm SFD}}(H_{{\rm r}})=\int_{-H_{{\rm s}}}^{H_{{\rm r}}}dH\,\rho(H,H_{{\rm r}})=-\frac{1}{2}\frac{dM(H,H_{{\rm r}})}{dH}\label{eq:SFD_Form}
\end{equation}
Both $\mathcal{A}$ and $\mathcal{F}$ features manifest as prominent peaks in the SFD (\ref{fig:FORC-Expts}(b)), which incidentally facilitates direct comparison with imaging results.

\begin{figure}
\noindent \begin{centering}
\includegraphics[width=3.3in]{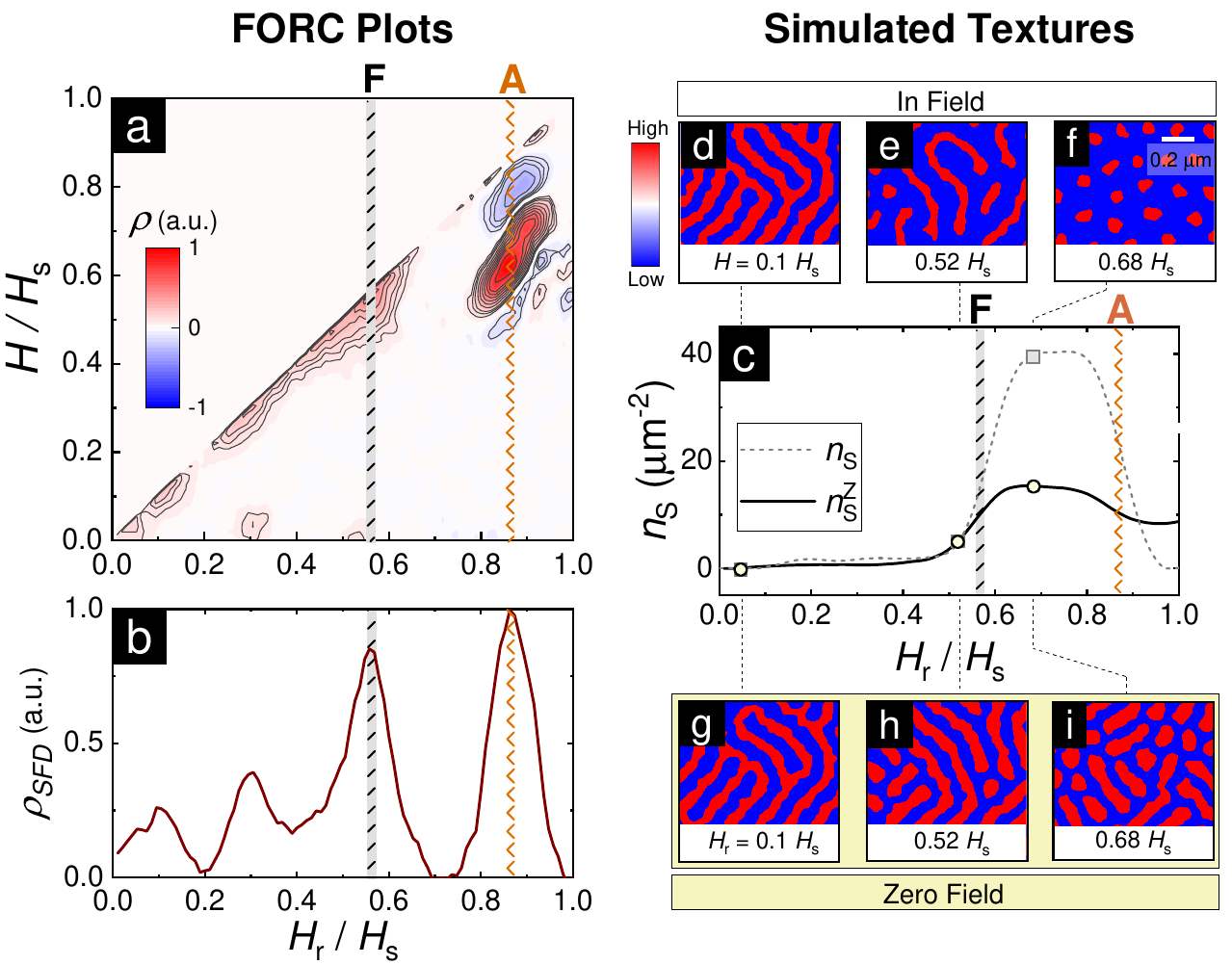}
\par\end{centering}
\noindent \caption[Simulated FORC \& ZF Textures]{\textbf{Simulated FORC Irreversibility and ZF Textures for Fe(4)/Co(6) stacks.} \textbf{(a)} Color plot of simulated FORC irreversibility $\rho(H,H_{{\rm r}})$, and  \textbf{(b)} the corresponding SFD obtained from micromagnetic simulations of FORCs for Fe(4)/Co(6) parameters. As in \ref{fig:FORC-Expts}, the two most prominent features are labeled as $\mathcal{A}$ and $\mathcal{F}$ respectively. \textbf{(c)} Corresponding plots of skyrmion densities of simulated IF and ZF magnetization with varying $H$ and $H_{{\rm r}}$ respectively. The $\mathcal{A}$ and $\mathcal{F}$ lines from (a-b) are overlaid for comparison. \textbf{(d-i)} Representative simulated magnetization images of (d-f) IF and (g-i) ZF morphology for selected $(H,H_{{\rm r}})$ pairs (corresponding to data markers in (c)). \label{fig:FORC-Sims}}
\end{figure}

\paragraph{IF \& ZF Densities}
\ref{fig:FORC-Expts}(c-i) detail the spin texture morphology of Fe(4)/Co(6) as imaged by MFM under IF and ZF conditions. In line with results reported on similar samples \citep{Soumyanarayanan2017}, IF textures evolve progressively from stripes to skyrmions (\ref{fig:FORC-Expts}(d-f)). The corresponding ZF morphology is remarkably non-monotonic with varying $H_{{\rm r}}$ -- showing skyrmions and short stripes at intermediate values (\ref{fig:FORC-Expts}(h-i), \ref{fig:FORC-Motivation}(c)), and labyrinthine state on either side (\ref{fig:FORC-Expts}(g), \ref{fig:FORC-Motivation}(e)). Notably, \ref{fig:FORC-Expts}(c) showing the IF and ZF skyrmion densities, i.e. $n_{{\rm S}}(H)$ and $n_{{\rm S}}^{Z}(H_{{\rm r}})$ respectively (identification protocols in \citep{SuM}) -- reveals several trends of interest. First, both $n_{{\rm S}}(H)$ and $n_{{\rm S}}^{Z}(H_{{\rm r}})$ display dome shape trends that peak $\sim0.7\,H_{{\rm s}}$ ($n_{{\rm S}}\sim40$~\textmu m$^{-2}$ and $n_{{\rm S}}^{Z}\sim12$~\textmu m$^{-2}$, \ref{fig:FORC-Expts}(c)). Next, a comparison with the FORC data (\ref{fig:FORC-Expts}(a-b)) shows that the $\mathcal{A}$ feature corresponds to a sharp drop in $n_{{\rm S}}$ -- as expected near annihilation. Crucially, the FORC $\mathcal{F}$ peak witnesses a sharp rise in both $n_{{\rm S}}(H)$ and $n_{{\rm S}}^{Z}(H_{{\rm r}})$. The coincidence of these features suggests that the $\mathcal{F}$ peak may be associated with IF skyrmion formation, and their ZF persistence.

\paragraph{Simulation Results}
We now turn to micromagnetic simulations of Fe(4)/Co(6) hysteresis loops, performed for a \emph{grain-free} environment, using FORC protocols to elucidate the relationship between the evolution of spin textures and their FORC signatures. The simulation results are shown in \ref{fig:FORC-Sims} (hysteresis curves in \citep{SuM}) to facilitate direct comparison with experiments (\ref{fig:FORC-Expts}). On one hand, the field evolution of simulated IF (\ref{fig:FORC-Sims}d-f) and ZF (\ref{fig:FORC-Sims}(g-i)) textures and the dome-shape trend in skyrmion densities (\ref{fig:FORC-Sims}(c)) are consistent with MFM experiments, with $n_{{\rm S}}^{Z}$ peaking over $0.6-0.8\,H_{{\rm s}}$. The domain walls consistently exhibit N\'{e}el helicity, and layer-wise chirality variations reported in previous studies \citep{Legrand2018,Dovzhenko2018} are not expected to play a significant role in this work \citep{SuM}. On the other hand, the simulated FORC irreversibility -- plotted as $\rho(H,H_{{\rm r}})$ (\ref{fig:FORC-Sims}(a)) and SFD (\ref{fig:FORC-Sims}(b)) -- also reproduces the key features seen in FORC experiments at comparable fields (\ref{fig:FORC-Expts}(a-b)). While the $\mathcal{A}$ peak ($\sim0.9\,H_{{\rm s}}$) coincides with a sharp drop in $n_{{\rm S}}$, the $\mathcal{F}$ peak ($\sim0.6\,H_{{\rm s}}$) correspondingly sees a sharp rise in $n_{{\rm S}}$. In addition to $\mathcal{A}$ and $\mathcal{F}$, the simulations show two smaller features over $0.1-0.4\,H_{{\rm s}}$, which are diminished in experiments. Overall the validation of key microscopic and macroscopic trends between \ref{fig:FORC-Sims} and \ref{fig:FORC-Expts} enable us to subsequently utilize the simulations to unravel the magnetic irreversibility encoded in the $\mathcal{F}$ feature.

\section{Microscopic Origin of Irreversibility\label{sec:Irrev-Microscopics}}

\begin{figure}
\noindent \begin{centering}
\includegraphics[width=3.3in]{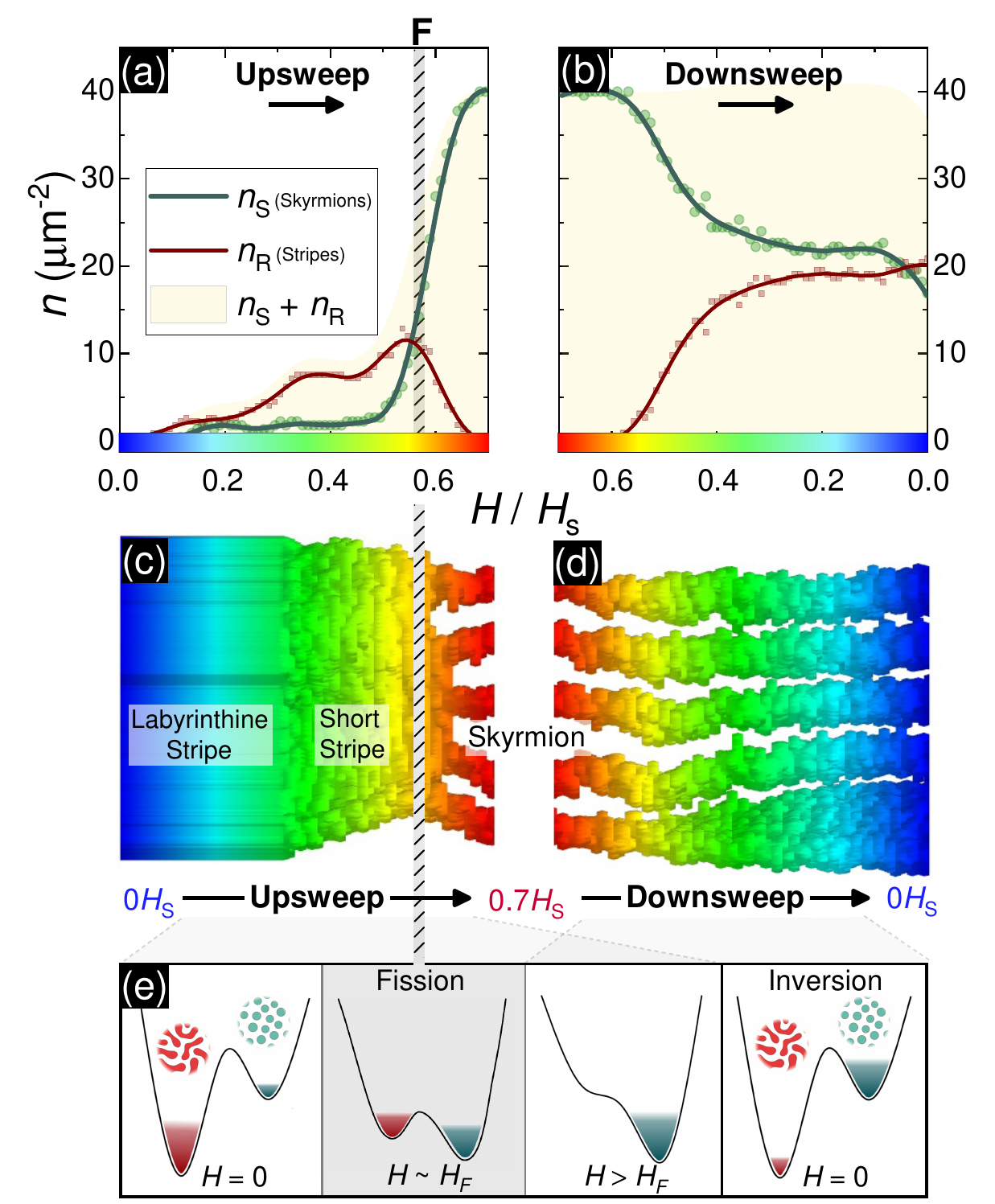}
\par\end{centering}
\noindent \caption[Microscopic Mechanism of Irreversibility]{\textbf{Microscopic Mechanism of Irreversibility from Fe(4)/Co(6) Simulations.} Microscopic analysis of the simulated Fe(4)/Co(6) FORC loop with $H_{{\rm r}}=0.7\,H_{{\rm s}}$ to elucidate the irreversibility of the stripe-skyrmion transition. \textbf{(a-b)} The field evolution of the density of skyrmions ($n_{{\rm S}}$, red), stripes ($n_{{\rm R}}$, black), and their sum total ($n_{{\rm tot}}$, shaded yellow) is shown on the upsweep ((a), $0\rightarrow0.7\,H_{{\rm s}}$) and downsweep ((b), $0.7\,H_{{\rm s}}\rightarrow0$) following negative saturation. A sharp change is seen at the guide line corresponding to $\mathcal{F}$ (from \ref{fig:FORC-Sims}(a-b)). \textbf{(c-d)} The field evolution of a prototypical stripe is shown along the FORC (c) upsweep and (d) downsweep, after its separation from the LS state (colorscale indicates field magnitude). The texture is identified and isolated for each field slice and stacked horizontally for direct comparison with (a-b). \textbf{(e)} Schematic depiction of the energetics of textures along the FORC loop. On the upsweep, stripes fission into skyrmions near $H_{\mathcal{F}}$ ($\mathcal{F}$ peak). Upon field reversal, the enhanced metastability of skyrmions enables their persistence to ZF (details in text). \label{fig:Irrev-Microscopics}}
\end{figure}

\paragraph{Field Evolution}
\noindent We elucidate the microscopic mechanism underlying the FORC irreversibility by examining the field evolution of simulated Fe(4)/Co(6) textures along the $H_{{\rm r}}\simeq0.7\,H_{{\rm s}}$ FORC loop. \ref{fig:Irrev-Microscopics}(a) shows the density evolution of stripes ($n_{{\rm R}}$) and skyrmions ($n_{{\rm S}}$) as the field is swept up from zero. To begin with, both $n_{{\rm R}}$ and $n_{{\rm S}}$ rise monotonically, and are comparable at $\sim0.6\,H_{{\rm s}}$ ($\mathcal{F}$). Beyond $\mathcal{F}$, $n_{{\rm R}}$ drops sharply to zero, while $n_{{\rm S}}$ concomitantly rises by 4-fold to reach its peak value at $\sim0.7\,H_{{\rm s}}$ -- indicating the fission of individual stripes into multiple skyrmions. Meanwhile, the down sweep from $H_{{\rm r}}\simeq0.7\,H_{{\rm s}}$ presents a contrasting picture of texture evolution (\ref{fig:Irrev-Microscopics}(b)). As $H$ is reduced, $n_{{\rm S}}$ gradually drops to half its maximal value, while $n_{{\rm R}}$ correspondingly rises from zero by a similar magnitude. Thus the total texture density, $n_{{\rm tot}}=n_{{\rm S}}+n_{{\rm R}}$ (shaded yellow in \ref{fig:Irrev-Microscopics}(a-b)) is nearly constant for the down-sweep, while increasing sharply after $\mathcal{F}$ for the up-sweep.

\begin{figure*}
\noindent \begin{centering}
\includegraphics[width=6in]{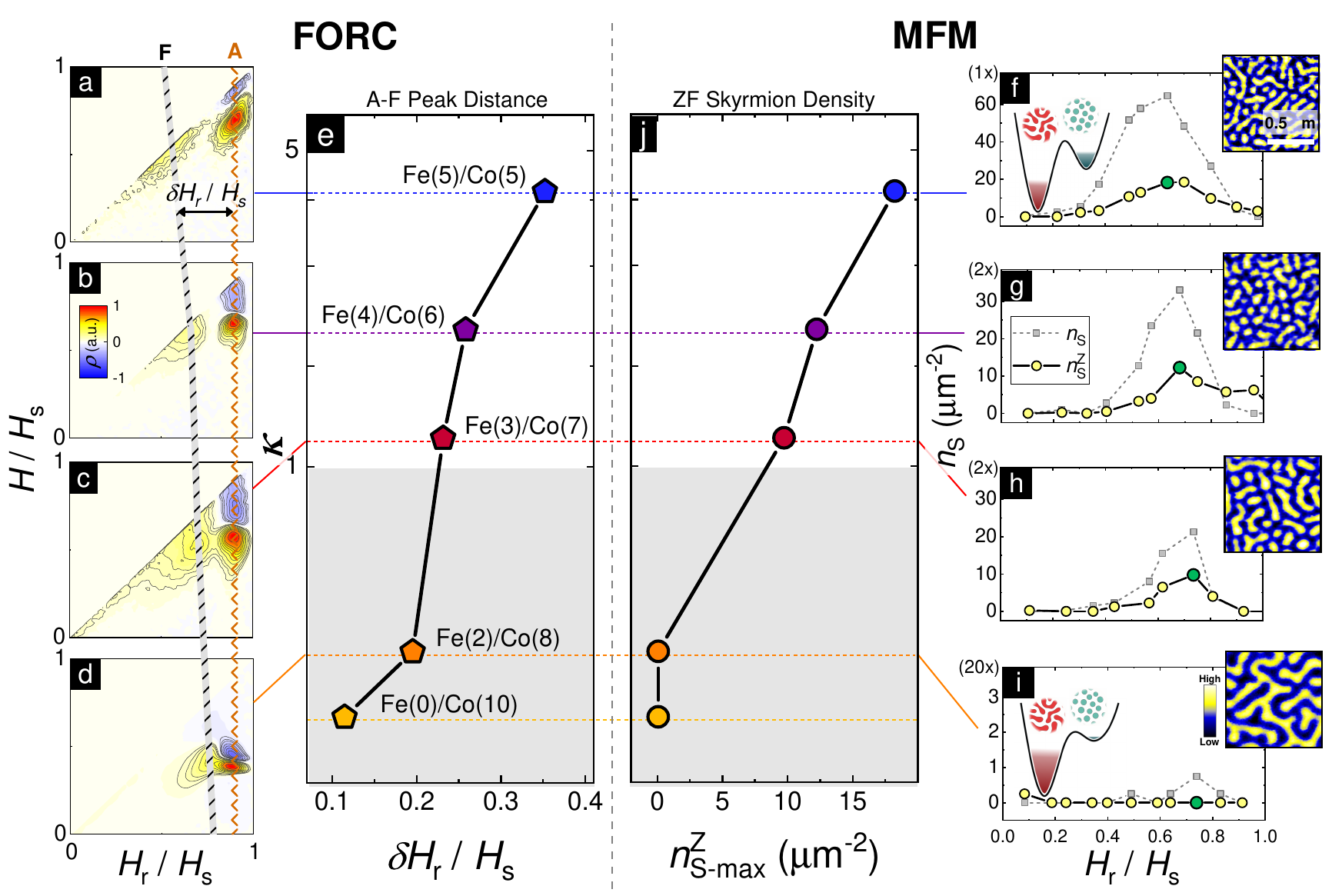}
\par\end{centering}
\noindent \caption[Thermodynamic Tuning of Irreversibility]{\textbf{Experimental Stability Parameter ($\kappa$) Modulation of Fission Process. }Experimentally measured variation of FORC irreversibility and ZF textures for the five Fe($x$)/Co($y$) samples investigated here, wherein $\kappa$ varies over $0.3-4.5$. \textbf{(a-d)} Color plots of FORC irreversibility, $\rho(H,H_{{\rm r}})$ (c.f. \ref{fig:FORC-Expts}a) for four samples corresponding to $\kappa$ values shown in (e). Dashed black \& orange lines indicate $\mathcal{F}$ and $\mathcal{A}$ peaks, which diverge with increasing $\kappa$. \textbf{(e)} Normalized distance, $\delta H_{{\rm r}}/H_{{\rm s}}$ , between $\mathcal{A}$ and $\mathcal{F}$ peaks plotted against $\kappa$, showing a monotonic increase. \textbf{(f-i)} MFM field evolution of skyrmion densities for IF($n_{{\rm S}}(H)$) and ZF ($n_{{\rm S}}^{Z}(H_{{\rm r}})$) configurations for the four samples examined in (a-d), showing dome-shaped field trends. Top right inset for each panel shows the ZF MFM image corresponding to the maximum in $n_{{\rm S}}^{Z}$ (green circle in $n_{{\rm S}}^{Z}(H_{{\rm r}})$). Left insets in (f),(i) illustrates the expected ZF energy landscape for samples with high and low $\kappa$ respectively. \textbf{(j)} The maximum ZF skyrmion density, $n_{{\rm S,}{\rm max}}^{Z}$ for all five samples plotted against $\kappa$. \label{fig:Irrev-SampleDep}}
\end{figure*}

\paragraph{Microscopic Irreversibility}
This hysteretic asymmetry is further illustrated in \ref{fig:Irrev-Microscopics}(c) (details in \citep{SuM}), which maps the FORC evolution of a prototypical simulated Fe(4)/Co(6) stripe following its separation from the labyrinthine state ($\sim0.3\,H_{{\rm s}}$). On the up sweep (\ref{fig:Irrev-Microscopics}(a): left), the stripe retains its morphology up to $\sim0.6\,H_{{\rm s}}$, after which it sharply fissions into five skyrmions. Notably, stripes are seldom found to shrink into individual skyrmions for Fe(4)/Co(6) parameters. Meanwhile on the down-sweep (\ref{fig:Irrev-Microscopics}(d): right), these skyrmions either retain their morphology down to ZF, albeit with $\sim$30\% increased size, or elongate to individual stripes, while preserving the total texture density. Crucially, we have never observed the merger of multiple skyrmions into one stripe.

\paragraph{Skyrmion/Stripe Asymmetry}
These microscopic trends establish a fundamental asymmetry in the field evolution of Fe(4)/Co(6) textures. While a stripe fissions into multiple skyrmions upon increasing field, these skyrmions do not fuse back into one stripe upon reducing field -- and often retain their morphology at ZF. The fission of individual stripes into multiple skyrmions near $\sim0.6\,H_{{\rm s}}$, is therefore irreversible, and manifests as the $\mathcal{F}$ peak in FORC experiments (\ref{fig:FORC-Expts}(c)). Moreover, this stark asymmetry in stripe-skyrmion transformation provides two valuable thermodynamic insights. First, it elucidates the energetics of skyrmion formation -- depicted schematically in \ref{fig:Irrev-Microscopics}(d). Stripes, while stable at lower fields, transform into skyrmions which are energetically preferred at higher fields. Upon field reversal, the metastability of skyrmions enables a sizable fraction to persist till ZF. Second, a likely origin of the enhanced metastability of skyrmions may be their morphology, as other energetic considerations (e.g. topological charge \citep{Bourianoff2018,Bernand-Mantel2017}) suggest that a stripe is equally likely to shrink to one skyrmion. Instead, the observed fission of stripes into multiple skyrmions indicates sizable entropic contributions to skyrmion stability, as independently suggested by theoretical works \citep{Hagemeister2015,Desplat2018}.

\section{Role of Thermodynamic Stability\label{sec:Irrev-SampleDep}}

\paragraph{Stack Tunability}
The thermodynamic factors underlying the observed persistence of Fe(4)/Co(6) skyrmions to ZF prompt us to investigate the role of the stability parameter $\kappa$ - which can be uniquely varied on either side of unity within Ir/Fe/Co/Pt multilayers \citep{Soumyanarayanan2017}. Accordingly, the FORC (\ref{fig:Irrev-SampleDep}(a-e)) and MFM (\ref{fig:Irrev-SampleDep}(f-j)) experiments, shown for Fe(4)/Co(6) in \ref{fig:FORC-Expts}, were also performed on four other samples, with $\kappa$ varied systematically over $0.3-4.1$. %

\paragraph{Sk Density Trends}
First, MFM imaging (\ref{fig:Irrev-SampleDep}(f-i)) shows that $n_{{\rm S}}(H)$ consistently increases in magnitude with $\kappa$, and exhibits a dome-shaped field trend in line with previous reports \citep{Soumyanarayanan2017}. For $\kappa>1$, the ZF density $n_{{\rm S}}^{Z}(H_{{\rm r}})$ increases with $\kappa$, reaching as high as 20~$\mu$m$^{-2}$ for Fe(5)/Co(5) (\ref{fig:Irrev-SampleDep}(f)). In this case, the $n_{{\rm S}}^{Z}(H_{{\rm r}})$ trend is much in line with $n_{{\rm S}}(H)$, as previously noted for Fe(4)/Co(6) (\ref{fig:FORC-Expts}). In contrast, for $\kappa<1$, $n_{{\rm S}}(H)$ is considerably lower, and peaks at fields closer to $H_{{\rm s}}$ (\ref{fig:Irrev-SampleDep}(i)). Crucially the ZF density $n_{{\rm S}}^{Z}(H_{{\rm r}})$ is negligible or zero for $\kappa<1$ (\ref{fig:Irrev-SampleDep}(i,j)). These trends, summarized in \ref{fig:Irrev-SampleDep}(j), could be explained by the enhanced metastability of skyrmions underlying their ZF persistence manifesting only for $\kappa>1$.

\paragraph{FORC Convergence}
Corresponding FORC measurements across $\kappa$, shown as $\rho(H,H_{{\rm r}})$ color plots (\ref{fig:Irrev-SampleDep}(a-d)), present a revealing thermodynamic picture. For $\kappa>1$, the $\mathcal{A}$ and $\mathcal{F}$ features are well-separated, and the $\mathcal{F}$-peak is pinned to the $H=H_{{\rm r}}$ edge (\ref{fig:Irrev-SampleDep}(a,b)). As $\kappa$ is reduced to unity, the $\mathcal{F}$-peak shifts from the edge and moves towards $\mathcal{A}$ (\ref{fig:Irrev-SampleDep}(c)), mirroring the observed $n_{{\rm S}}(H)$ trend across samples. Notably, for $\kappa<1$ the $\mathcal{A}$ and $\mathcal{F}$ features are merged (\ref{fig:Irrev-SampleDep}(d)), despite the presence of skyrmions at finite fields. This suggests that fission, the presence of which is indicated by the $\mathcal{F}$ feature, is not the only skyrmion formation mechanism, and incidentally explains the absence of the latter in FORC studies of stripe-bubble systems\citep{Davies2004,Davies2008,Kirby2010}. Finally, the $\kappa$-dependence of the $\mathcal{A}$-$\mathcal{F}$ separation ($\delta H_{{\rm r}}/H_{{\rm s}}$, \ref{fig:Irrev-SampleDep}(e)) sheds light on the $n_{{\rm S}}^{Z}$ trends detailed above (\ref{fig:Irrev-SampleDep}(j)). Increasing $\kappa$ enhances the separation between fission and annihilation processes, resulting in an annihilation-free environment conducive to enhanced skyrmion nucleation and stability. Conversely, smaller $\mathcal{A}$-$\mathcal{F}$ separation may result in overlap of skyrmion formation and annihilation regimes, thereby lowering the skyrmion density. Such direct correspondence between textural transitions and their FORC signatures provides a high-throughput predictive capability to engineer topological spin textures with enhanced stability.%

\section{Summary \& Impact\label{sec:Discussion}}

\paragraph{Insights Summary}
While FORC techniques have previously been used to elucidate domain nucleation and annihilation \citep{Davies2004,Davies2008,Kirby2010} and control their morphology \citep{Westover2016,Fallarino2019,Desautels2019}, our work extends this toolkit to unveil a topological transition between spin textures in $\kappa>1$ multilayers. At intermediate magnetic fields en route to saturation, individual stripes in these samples are shown to irreversibly fission into multiple skyrmions -- exhibiting a unique FORC irreversibility signature. Crucially, the skyrmions thus formed are robust to field reversal, with many retaining their morphology down to ZF. Finally, reducing $\kappa$ smoothly diminishes both the fission process and the ZF skyrmion density -- both are absent for $\kappa<1$. The microscopic and thermodynamic characteristics of this transition -- established by combining imaging, magnetometry, and simulations -- have profound implications on future directions concerning the energetics and ambient stability of multilayer skyrmions.

\paragraph{Energetics Impact}
Our results shed light on the energetic barrier separating skyrmions and stripes -- chiral magnetic textures of equal topology but different morphology -- in multilayer films \citep{Singh2019,Lemesh2018}. Theoretical works on skyrmion stability have investigated skyrmion energetics with respect to a uniformly magnetized state by modeling it as a single circular magnetic domain of varying radius \citep{Kiselev2011,Rohart2013,Leonov2016,Bernand-Mantel2017,Buttner2018}. Such analytical models can predict skyrmion stability with respect to the uniform state, especially for low DMI ($\kappa<1$) samples. However, they have limited utility in examining more complex magnetic evolutions, such as transitions between a stripe and a skyrmion, or a stripe and multiple skyrmions. Our work suggests that transitions involving multiple spin textures increase in significance for high DMI samples ($\kappa>1$) -- especially as we approach the ZF limit relevant to practical applications. In these cases, the energetics governing skyrmion formation and decay should involve sizable morphological and entropic components. We posit that these additional considerations should be incorporated into future theoretical frameworks on the energetics of skyrmion evolution.

\paragraph{Materials Impact}
Crucially, our work provides immediate, practical directions on materials and techniques to achieve stable multilayer skyrmions under ambient conditions. First, FORC characterization - herein established as a thermodynamic marker of skyrmion creation - can serve to enhance the throughput of skyrmion materials development. Notably, the proximity of the FORC fission process ($\mathcal{F}$-peak) to the $H-H_{{\rm r}}=0$ edge is expected to stabilize ZF skyrmions. This consideration, in conjunction with the magnitude of $\kappa$, are strong indicators of ZF skyrmion stability in chiral multilayers. Next, our work motivates the use of minor loops and current-induced Oersted field effects - previously shown to achieve domain nucleation \citep{Davies2004,Berger2010,Woo2016,Westover2016,Fallarino2019} -- towards manipulating chiral spin textures. In this case, the Oersted field magnitude and repetition rate could serve as ``knobs'' to control, for example, the ZF texture density in devices. Specifically, Oersted field manipulation techniques could be extended to achieve reshufflers \citep{Pinna2018} or to construct weighting functions towards recursive neural networks \citep{Prychynenko2018} -- relevant to applications in brain-inspired computing. 

\noindent \begin{center}
{\small{}\rule[0.5ex]{0.4\columnwidth}{0.5pt}}{\small\par}
\par\end{center}

\noindent {\small{}We acknowledge the support of the National Supercomputing
Centre (NSCC) for computational resources. This work was supported
by the SpOT-LITE programme (Grants No. A1818g0042 and No. A18A6b0057), funded
by Singapore's RIE2020 initiatives, and by the Pharos Skyrmion programme
(Grant No. 1527400026) funded by A{*}STAR, Singapore.}\textsf{\textbf{\small{}}}%

\noindent \begin{center}
{\small{}\rule[0.5ex]{0.4\columnwidth}{0.5pt}}{\small\par}
\par\end{center}

\newpage
\widetext
\setcounter{secnumdepth}{4}
\setcounter{section}{0}
\setcounter{equation}{0}
\setcounter{figure}{0}
\setcounter{table}{0}
\setcounter{page}{1}

\makeatletter
\renewcommand\thesection{S\arabic{section}}
\renewcommand{\theparagraph}{S\arabic{section}\alph{paragraph}}
\makeatletter\@addtoreset{paragraph}{section}\makeatother
\makeatletter\def\p@paragraph{}\makeatother
\renewcommand{\thefigure}{S\arabic{figure}}
\renewcommand{\theequation}{S\arabic{equation}}
\renewcommand{\thetable}{S\arabic{table}}
\addto\captionsenglish{\renewcommand{\figurename}{FIG.}}

\setcounter{tocdepth}{1}
\makeatletter\def\l@section{\@dottedtocline{1}{0.6em}{1.5em}}\makeatother
\makeatletter\def\l@paragraph{\@dottedtocline{4}{1.5em}{1.8em}}\makeatother
\makeatletter\def\l@figure{\@dottedtocline{1}{0.6em}{1.8em}}\makeatother

\linespread{1.25}
\def\arraystretch{1.5}
\setlength{\parskip}{1.5ex plus0.2ex minus0.2ex} 
\setlength{\abovecaptionskip}{4pt}\setlength{\belowcaptionskip}{-4pt}
\setlength{\abovedisplayskip}{0ex}\setlength{\belowdisplayskip}{0ex}
\setlength{\abovedisplayshortskip}{0ex}\setlength{\belowdisplayshortskip}{0ex}

\titleformat{\section}{\large\bfseries\scshape\filcenter}{\thesection.}{1em}{#1}[{\titlerule[0.5pt]}]
\titlespacing*{\section}{0pt}{1ex}{1ex}
\titleformat{\subsection}{\bfseries\sffamily}{\thessubsection.}{0em}{#1}
\titlespacing{\subsection}{0pt}{0.5ex}{0.5ex}
\titleformat{\paragraph}[runin]{\sffamily\bfseries}{}{-1.2em}{#1.}
\titlespacing*{\paragraph}{1.25em}{2ex}{0.4em}[] 

\begin{center}
	\Large\textbf{Supplementary Materials for\\}
	\large\textbf{Skyrmion Generation from Irreversible Fission of Stripes in Chiral Multilayer Films\smallskip{}}
\end{center}

\linespread{2}
\setlength{\parskip}{3ex plus0.2ex minus0.2ex}

\linespread{1.25}
\setlength{\parskip}{1ex plus0.2ex minus0.2ex}

\noindent \section{Stack Structure \& Magnetic Properties\label{sec:MagProps}}

\begin{figure}[h]
\begin{centering}
\includegraphics[width=2.5in]{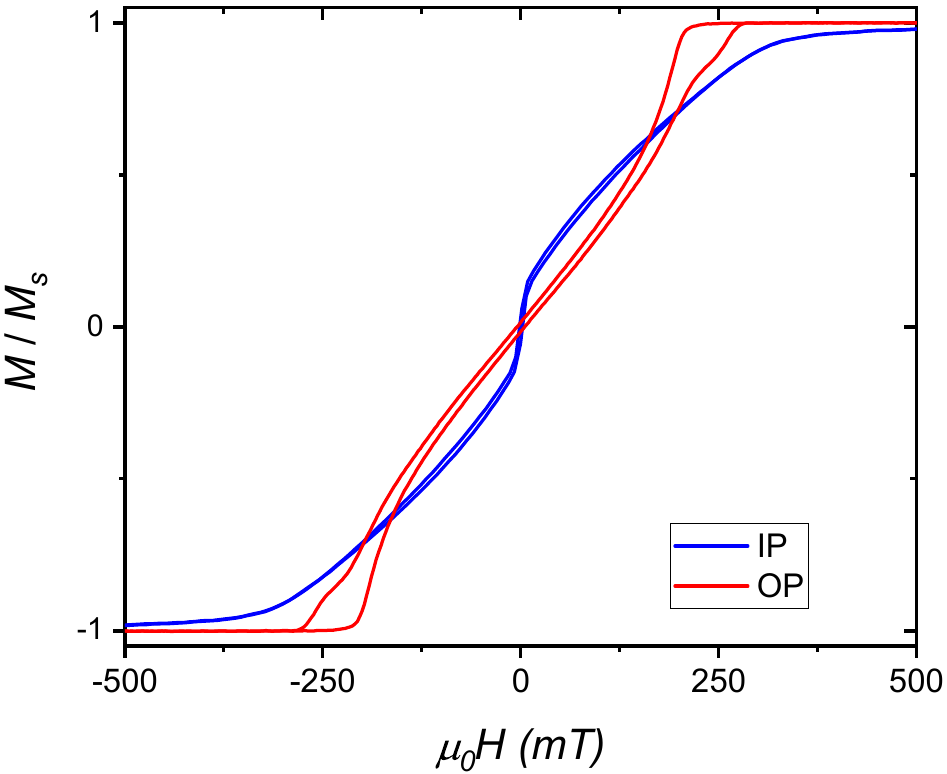}
\par\end{centering}
\noindent \caption[Hysteresis Loop Example]{\textbf{$M(H)$ Characterization. }Hysteresis loops of normalized magnetization, $M/M_{{\rm s}}$ for sample Fe(4)/Co(6) measured by alternating gradient magnetometry (AGM) in out-of-plane (OP, red) and in-plane (IP, blue) configurations.\textbf{ }\label{fig:ExptMHLoop}}
\end{figure}

\paragraph{Magnetic Properties}
\ref{tab:SamplesMagProps} shows the relevant magnetic parameters for the five {[}Ir/Fe($x$)/Co($y$)/Pt{]} sample compositions studied here. The saturation magnetization, $M_{{\rm s}}$ and the effective anisotropy, $K_{{\rm eff}}$ were determined from out-of-plane (OP) and in-plane (IP) magnetization loops (see e.g. \ref{fig:ExptMHLoop}) -- the latter from the areal difference between the two loops \citep{Johnson_1996}. Meanwhile, the OP saturation field, $H_{\rm s}$ -- determined from the FORC distribution, $\varrho(H,H_{\rm r})$ -- is defined as coinciding with the end of the domain annihilation process (top right of \ref{fig:AGM-FORC}c). Together with the measured zero field (ZF) domain periodicity, $P^{{\rm ZF}}$ and corresponding micromagnetic simulations, these were used to determine the exchange interaction, $A$ and Dzyaloshinskii-Moriya interaction, $D$, similar to previous work \citep{MoreauLuchaire2016,Woo2016,Soumyanarayanan2017,Ho2019}.

\noindent
\begin{table}[h]
\begin{tabular}{|c|c|c|c|c|c|c|c|}
\hline
\textbf{Sample } & $M_{{\rm s}}$  & $K_{{\rm eff}}$  & $P^{{\rm ZF}}$ & \textbf{$K_{{\rm u}}$}  & $D$  & \multicolumn{1}{c||}{$A$ } & \multirow{2}{*}{$\kappa$}\tabularnewline
\textbf{(14$\times$)} & (MA/m) & (mJ/m$^{3}$) & (nm) & (mJ/m$^{3}$) & (mJ/m$^{2}$) & \multicolumn{1}{c||}{(pJ/m)} & \tabularnewline
\hline
\hline
Fe(0)/Co(10) & \enskip{}1.16\enskip{} & \enskip{}0.60\enskip{} & \enskip{}346\enskip{} & \enskip{}1.45\enskip{} & \enskip{}0.9\enskip{} & \enskip{}17.8\enskip{} & \quad{}0.28\quad{} \tabularnewline
\hline
\multirow{1}{*}{Fe(2)/Co(8)} & 1.14 & 0.26 & 186 & 1.08 & 1.2 & 12.8 & 0.39\tabularnewline
\hline
\multirow{1}{*}{Fe(3)/Co(7)} & 1.11 & 0.10 & 152 & 0.87 & 1.7 & 13.2 & 1.15\tabularnewline
\hline
Fe(4)/Co(6) & 0.95 & 0.04 & 130 & 0.61 & 1.9 & 13.6 & 2.01\tabularnewline
\hline
Fe(5)/Co(5) & 1.06 & -0.01 & 110 & 0.69 & 2.1 & 13.6 & 4.06\tabularnewline
\hline
\end{tabular}\break\caption{\textbf{Magnetic Properties of {[}Ir/Fe($x$)/Co($y$)/Pt{]}$_{14}$
Multilayers.} The values of saturation magnetization ($M_{{\rm s}}$), effective anisotropy ($K_{{\rm eff}}$), zero field domain periodicity ($P^{{\rm ZF}}$ ), uniaxial anisotropy (\textbf{$K_{{\rm u}}$}), Dzyaloshinskii-Moriya interaction ($D$), exchange stiffness ($A$), and thermodynamic stability parameter ($\kappa$) for the five Fe($x$)/Co($y$) sample compositions studied in this work. \label{tab:SamplesMagProps}}
\end{table}

\clearpage{}

\section{Skyrmion Identification in MFM Images\label{sec:MFM}}

\begin{figure}[h]
    \centering
    \includegraphics[width=4in]{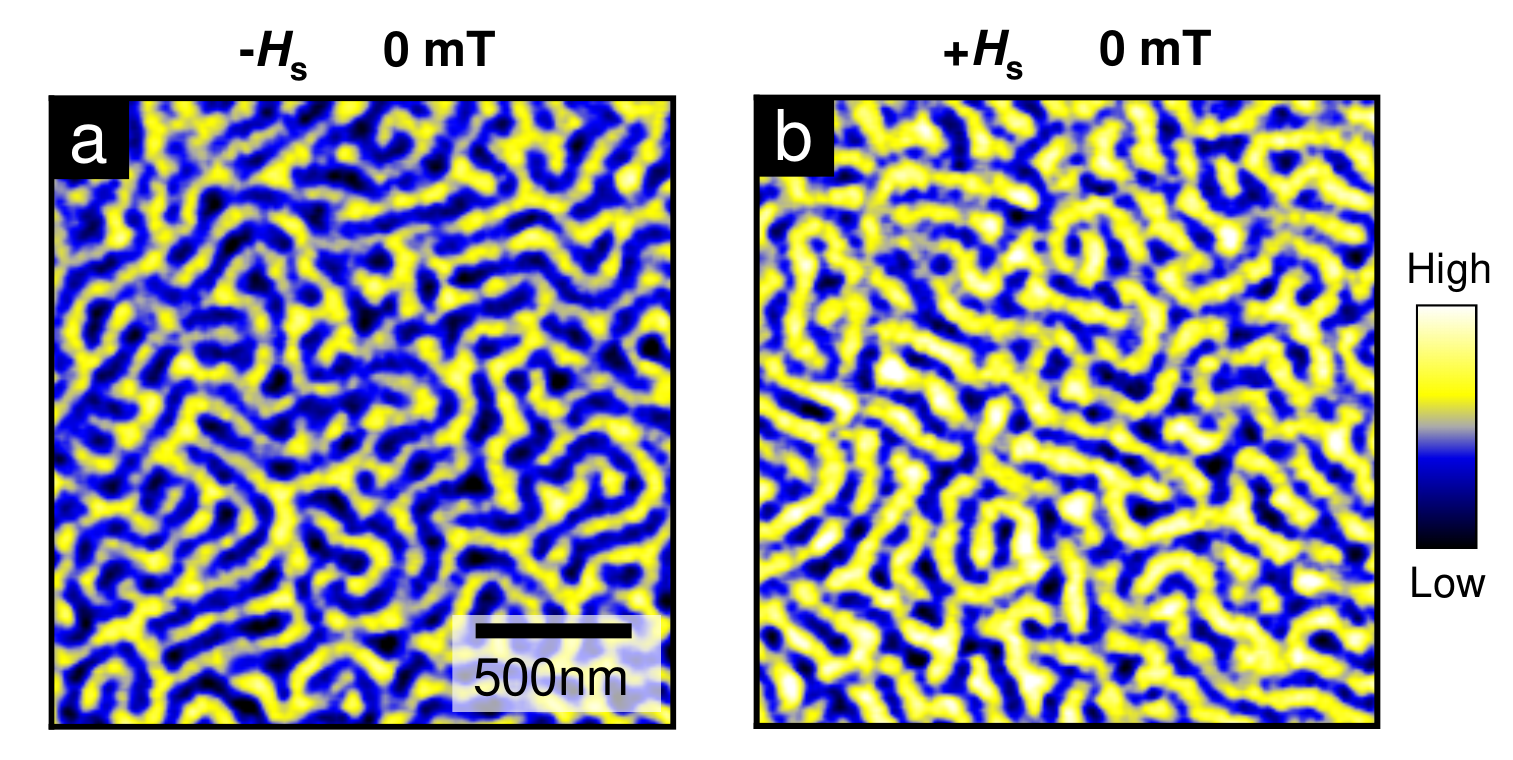}
    \caption{\textbf{Zero Field Domain Morphology.} MFM images of the ZF domain morphology for sample Fe(4)/Co(6) acquired following saturation at: (a) $H < -H_{\rm s}$ (large negative fields) and (b) $H > +H_{\rm s}$ (large positive fields).  }
    \label{fig:ZF_domain}
\end{figure}

\paragraph{Zero Field Domain Morphology}
The protocol followed for FORC and minor loop MFM experiments in this manuscript begins with saturating the sample at large, negative fields, i.e. $H < -H_{\rm s}$. To examine the generality of the protocol, we compare in \ref{fig:ZF_domain} the zero field (ZF) morphology obtained following saturation at large negative (\ref{fig:ZF_domain}a) and large positive (\ref{fig:ZF_domain}b) fields. Notably, a labyrinthine stripe morphology at ZF is obtained in both cases with similar lengthscales and periodicities. Meanwhile, the key observable difference is the inversion of MFM phase contrast between the two images, which may be explained by the flipping of remnant magnetization between the two saturation protocols. Overall, the consistency of ZF domain features between \ref{fig:ZF_domain}a-b suggest that the MFM imaging results reported in the manuscript are robust to systemic variations of the saturation protocol

\begin{figure}[h]
\begin{centering}
\includegraphics[width=5.1in]{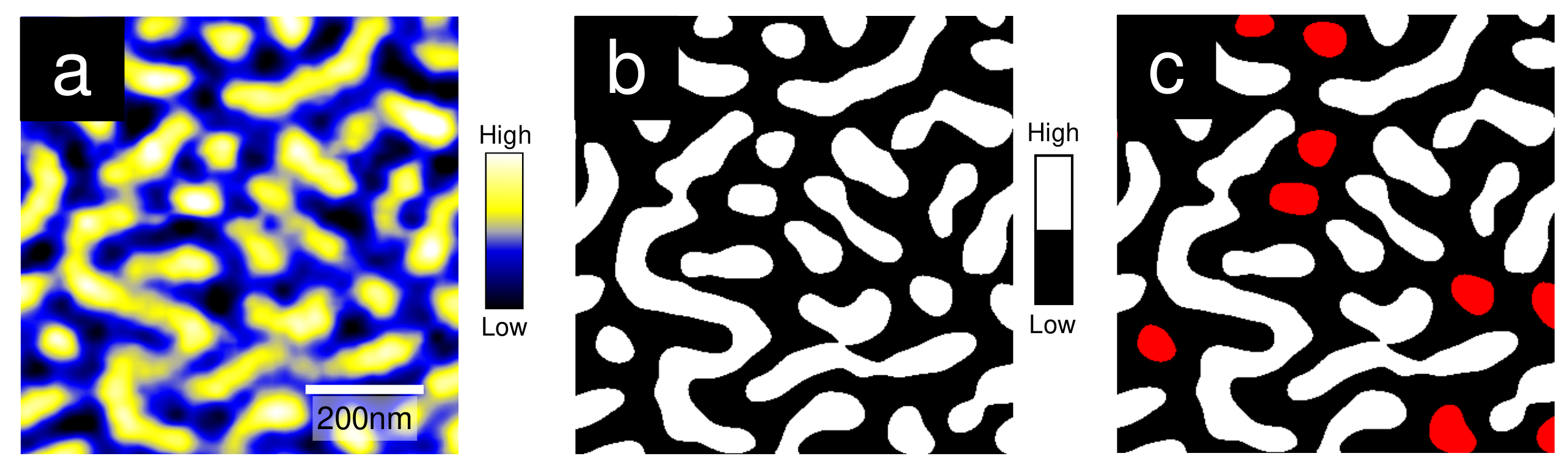}
\par\end{centering}
\noindent \caption[Skyrmion Identification Protocol]{\textbf{Protocol for Skyrmion Identification in Images.} \textbf{(a)} Post-processed MFM image showing typical magnetic features. \textbf{(b)} Binarized output from (a) following threshold segmentation. \textbf{(c)} Composite image of (b) showing clusters identified as skyrmions (in red).\textbf{ }\label{fig:MFM_SkID}}
\end{figure}

\paragraph{MFM Imaging of Skyrmions}
We have previously established that sub-100~nm magnetic textures observed in MFM images of Ir/Fe($x$)/Co($y$)/Pt multilayers are N\'{e}el-textured skyrmions \citep{Soumyanarayanan2017}. The size, profile, and periodicity of these skyrmions are demonstrably consistent across MFM and transmission XMCD microscopy techniques. Subsequent efforts on similar Ir/Fe/Co/Pt multilayers have also established the sensitivity of MFM to the texture, helicity and inhomogeneity of skyrmions \citep{Soumyanarayanan2017,Yagil2018}, and extended it over a wide range of temperatures \citep{Raju2019}. Here, we describe the protocol adopted to identify skyrmions in MFM images and characterize their densities at remanence, as well as finite fields.

\paragraph{Identification Protocol}
The MFM images acquired here were first processed with polynomial surface subtraction and low pass filtering -- commonly employed for scanning probe images (\ref{fig:MFM_SkID}a). Next, a segmentation process was used to separate two classes of pixels, i.e. binarize the image based on a threshold intensity value - obtained via Otsu's method \citep{Otsu1979}. The threshold value is optimized by maximizing the inter-class variance \citep{Otsu1979}. The binary clusters are then grouped into individual grains with connectivity of 4, and the grain properties -- diameter, $d_{{\rm S}}$, circularity, $C_{{\rm S}}$, and major/minor axes ratio, $r_{{\rm S}}$ -- are retrieved. The individual grains are then identified as a skyrmion based on these 3 indicators, with $d_{{\rm S}}<d_{{\rm th}}=150$~nm, $C_{{\rm S}}>C_{{\rm th}}=0.6$, $r_{{\rm S}}<r_{{\rm th}}=1.75$. These threshold values were established based on a large MFM dataset ($>1000$ skyrmions) with varying skyrmion sizes and densities, including results from our previous studies \citep{Soumyanarayanan2016}.

\clearpage{}

\section{FORC Measurements and Analysis\label{sec:AGM-FORC}}

\begin{figure}[h]
\begin{centering}
\includegraphics[width=6.5in]{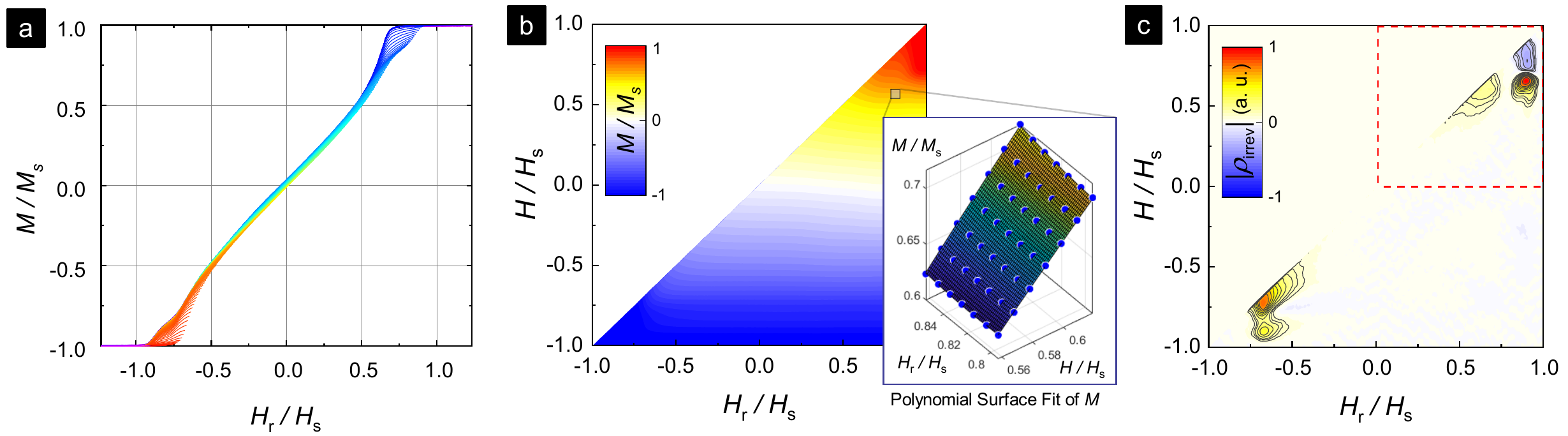}
\par\end{centering}
\noindent \caption[AGM \& FORC Measurements]{\textbf{AGM and FORC Measurements. (a)} Set of FORCs ($M(H)$) obtained for sample Fe(4)/Co(6) with $H_{{\rm r}}$, $H$ intervals of 3~mT. \textbf{(b)} Magnetization color plot of the FORC data shown in (a), projected on $H$, $H_{{\rm r}}$ axis. Zoomed inset shows the second order polynomial surface fit to $M(H,H_r)$ using \ref{eq:FORC_MHFit}. \textbf{(c)} Color plot of FORC irreversibility distribution, $\varrho(H,H_{{\rm r}})$, derived from the data shown in (a-b). Dashed box highlights the region of interest in $(H,H_{{\rm r}})$ -- shown in manuscript Figs. 2-5.\textbf{ }\label{fig:AGM-FORC}}
\end{figure}

\paragraph{FORC Measurements}
Prior to measuring a FORC, the sample is saturated at $-H_{0}=-500$~mT (over $1.5\,H_{{\rm s}}$). The field is subsequently increased to the targeted $H_{{\rm r}}$, and a FORC is traced out by measuring $M(H)$ as the field is decreased from $H_{{\rm r}}$ to $-H_{0}$. During the field reversal, magnetization measurements are recorded at field steps of 2-3~mT and averaged for 0.1~s per point. The processed is repeated to obtain a set of FORCs with $H_{{\rm r}}$ values ranging from $-1.5\,H_{{\rm s}}$ to $+1.5\,H_{{\rm s}}$ with intervals of 2-3~mT (\ref{fig:AGM-FORC}a-b). The dataset is further processed via Equation (1) (see main text) by fitting the magnetization to a second order polynomial surface \citep{Pike1999}\\
\begin{equation}
M(H,H_{{\rm r}})=a_{1}+a_{2}H_{{\rm r}}+a_{3}H+a_{4}H_{{\rm r}}^{2}+a_{5}H^{2}+a_{6}H_{{\rm r}}H\label{eq:FORC_MHFit}
\end{equation}
\\to produce a FORC distribution, $\varrho(H,H_{{\rm r}})$ (\ref{fig:AGM-FORC}c). The number of points used for fitting is determined by a smoothing factor, ${\rm SF}$, which is 3 in our work. The $\varrho(H,H_{{\rm r}})$ region of interest for our study is highlighted in \ref{fig:AGM-FORC}c (reproduced in manuscript Fig 2a), and corresponds to the positive quadrant, i.e. $H,H_{{\rm r}}>0$.

\begin{figure}[h]
\begin{centering}
\includegraphics[width=4.2in]{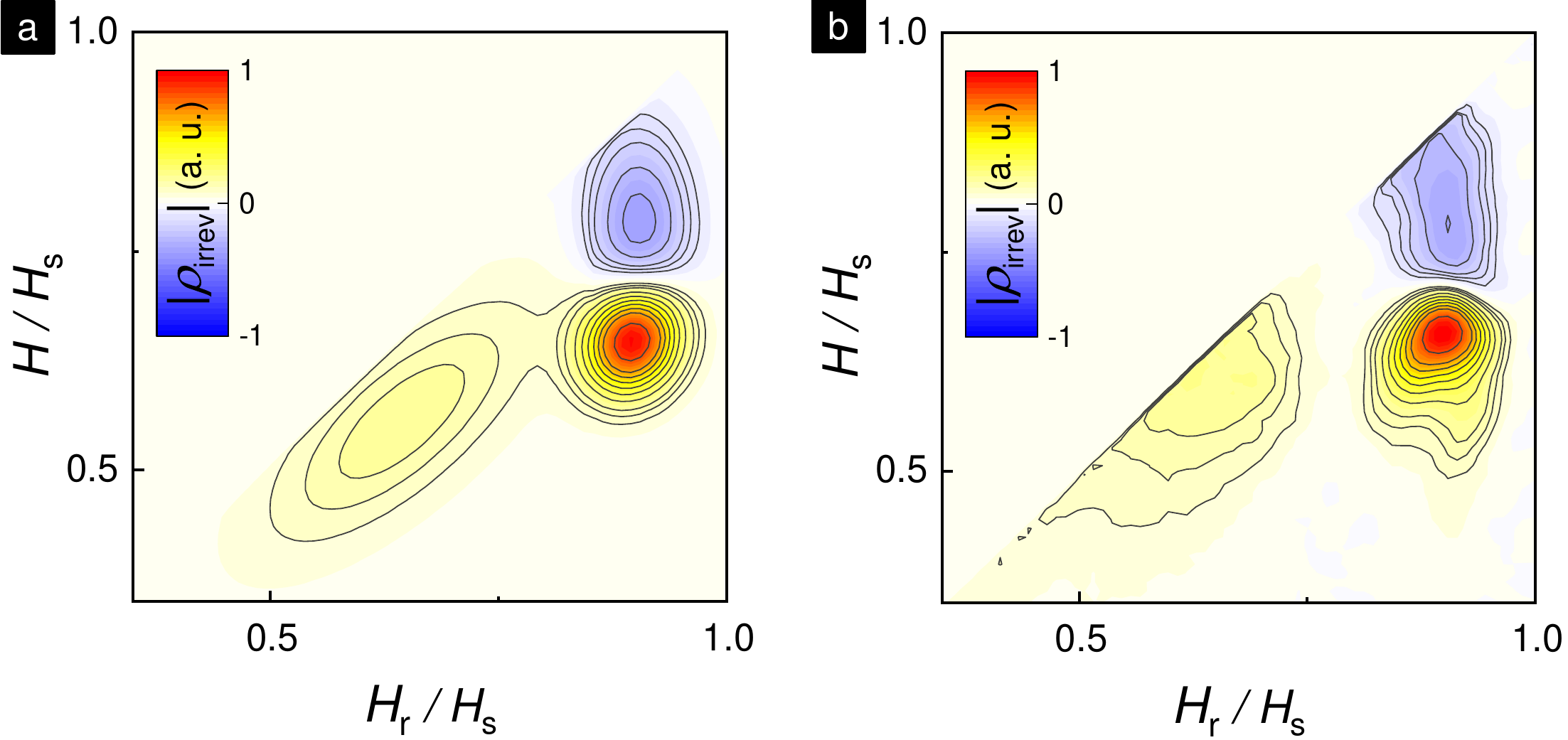}
\par\end{centering}
\noindent \caption[FORC Peak Fits]{\textbf{Fitting of $\mathcal{F}$ and $\mathcal{A}$ Features.} \textbf{(a)} Colour plot of fitted FORC $\varrho(H,H_{{\rm r}})$ of `F' and `A' features of \textbf{(b)} raw FORC $\varrho(H,H_{{\rm r}})$ (c.f. Fig. 2a) observed in Fe(4)/Co(6).\label{fig:FORC-PkFit}}
\end{figure}

\paragraph{FORC Peak Fitting Protocol}
Here, we describe the fitting procedure used to quantify the separation between the $\mathcal{A}$- and $\mathcal{F}$ peaks for manuscript Fig. 5. The peak features in the FORC distribution colour plot can be fitted using 2D Gaussian elliptical profiles. The general form of a sum of three 2D Gaussian elliptical profiles can be described by the equation:
\noindent
\begin{equation}
\varrho(H,H_{{\rm r}})=\sum_{i=1}^{3}A_{i}\,\exp\left[-\left\{ a_{i}(H-H_{i}^{0})^{2}+2b_{i}(H-H_{i}^{0})(H_{{\rm r}}-H_{i}^{1})+c_{i}(H_{{\rm r}}-H_{i}^{1})^{2}\right\} \right]+\,C\,,\label{eq:FORCPkFit}
\end{equation}
where $A_{i}$ are the peak amplitudes, $C$ is an overall offset, and $(H_{i}^{0},H_{i}^{1})$ are the peak centre coordinates of interest. The coefficients $(a_{i},b_{i},c_{i})$ describe the elliptical asymmetry of the peaks, and can be expressed as:\\
\begin{alignat}{2}
a_{i} & \:=\: & \frac{\cos^{2}\theta_{i}}{2\sigma_{i}^{2}}+\frac{\sin^{2}\theta_{i}}{2\omega_{i}^{2}}\,,\nonumber \\
b_{i} & \:=\: & \frac{\sin2\theta_{i}}{4\sigma_{i}^{2}}+\frac{\sin2\theta_{i}}{4\omega_{i}^{2}}\,,\nonumber \\
c_{i} & \:=\: & \frac{\sin^{2}\theta_{i}}{2\sigma_{i}^{2}}+\frac{\cos^{2}\theta_{i}}{2\omega_{i}^{2}}\,,\label{eq:FORCPkFit_Coeffs}
\end{alignat}
\\where $\sigma_{i}$ and $\omega_{i}$ are the spreads along $H$ and $H_{{\rm r}}$ respectively, and $\theta_{i}$ allows for the clockwise rotation of the ellipse with respect to the $H$-axis.

\paragraph{$\mathcal{A-\mathcal{F}}$ Peak Distance}
In our case, the $H_{{\rm r}}$ coordinates of the FORC peaks are determined from such Gaussian fits. For samples Fe(5)/Co(5) and Fe(4)/Co(6), the $\mathcal{F}$-peak is well-separated from the $\mathcal{A}$-peak, and so independent Gaussian fits are used for the 2 peaks. Meanwhile, for Fe(0)/Co(10), Fe(2)/Co(8), Fe(3)/Co(7), the $\mathcal{A}$- and $\mathcal{F}$-features are in closer proximity, and so two fitting approaches are investigated. First, all 3 peaks are fitted simultaneously using a 3-Gaussian function. Second, the pair of $\mathcal{A}$-peaks are fitted using a 2-Gaussian function, while the $\mathcal{F}$-peak is fitted independently. Both approaches yield similar values of peak separation. These results are shown in manuscript Fig. 5e in units of $\delta H_{{\rm r}}/H_{{\rm s}}$.

\clearpage{}

\noindent \section{Micromagnetic Simulations and Analysis\label{sec:SIMS}}

\begin{figure}[h]
\begin{centering}
\includegraphics[width=6.3in]{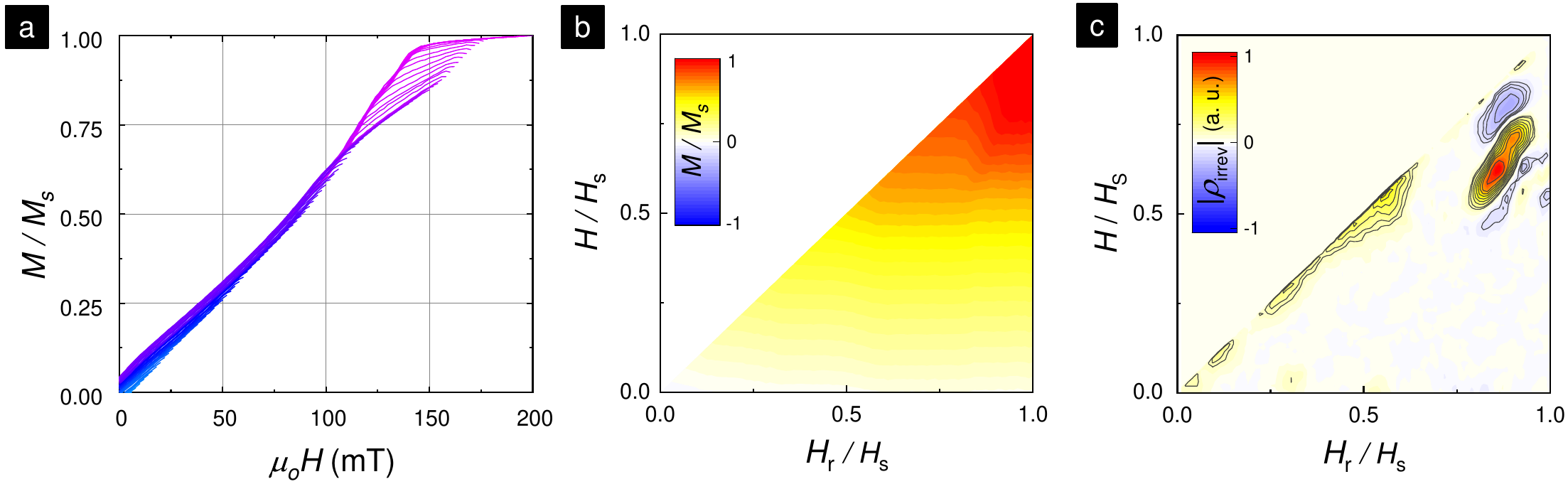}
\par\end{centering}
\noindent \caption[Simulated FORC Protocol]{\textbf{Simulated FORC Measurements.} \textbf{(a)} Simulated FORCs ($M(H)$) of sample Fe(4)/Co(6) with $H_{{\rm r}}$ and $H$ intervals of 2~mT. \textbf{(b) }Colour plot of simulated FORCs in (a), projected on $H$, $H_{{\rm r}}$ axis with colors indicating the magnetization. \textbf{(c) }Simulated FORC distribution, $\varrho(H,H_{{\rm r}})$, derived from set of FORCs in (a-b) using using manuscript Eqn 1.\textbf{ }\label{fig:Sims_FORC}}
\end{figure}

\paragraph{Simulated FORCs}
Micromagnetic simulations of FORC loops were performed by following the experimental FORC protocols as closely as possible. The finite temperature simulations -- implemented in mumax$^3$ by adding a randomly fluctuating Langevin term to the micromagnetic equation \citep{Vansteenkiste2014} -- implicitly incorporate entropic effects \citep{Brown_1993,Lyberatos_1993}. To begin with, a major hysteresis loop was simulated at 2~mT intervals, and the full spatial configuration of magnetization was recorded at each field -- to serve as inputs for the FORC simulations. Next, each FORC loop is simulated independently, starting from the magnetization configuration of the major loop ($H=H_{{\rm r}}$) and ending at ZF (\ref{fig:Sims_FORC}a-b). To reduce computation time, the simulations were performed only for the quadrant with $H,H_{{\rm r}}>0$ -- which contains the experimental features relevant to this work. The FORC analysis protocol used to produce \ref{fig:Sims_FORC}c followed the experimental procedures described in \ref{sec:AGM-FORC}.

\begin{figure}[h]
\begin{centering}
\includegraphics[width=4.0in]{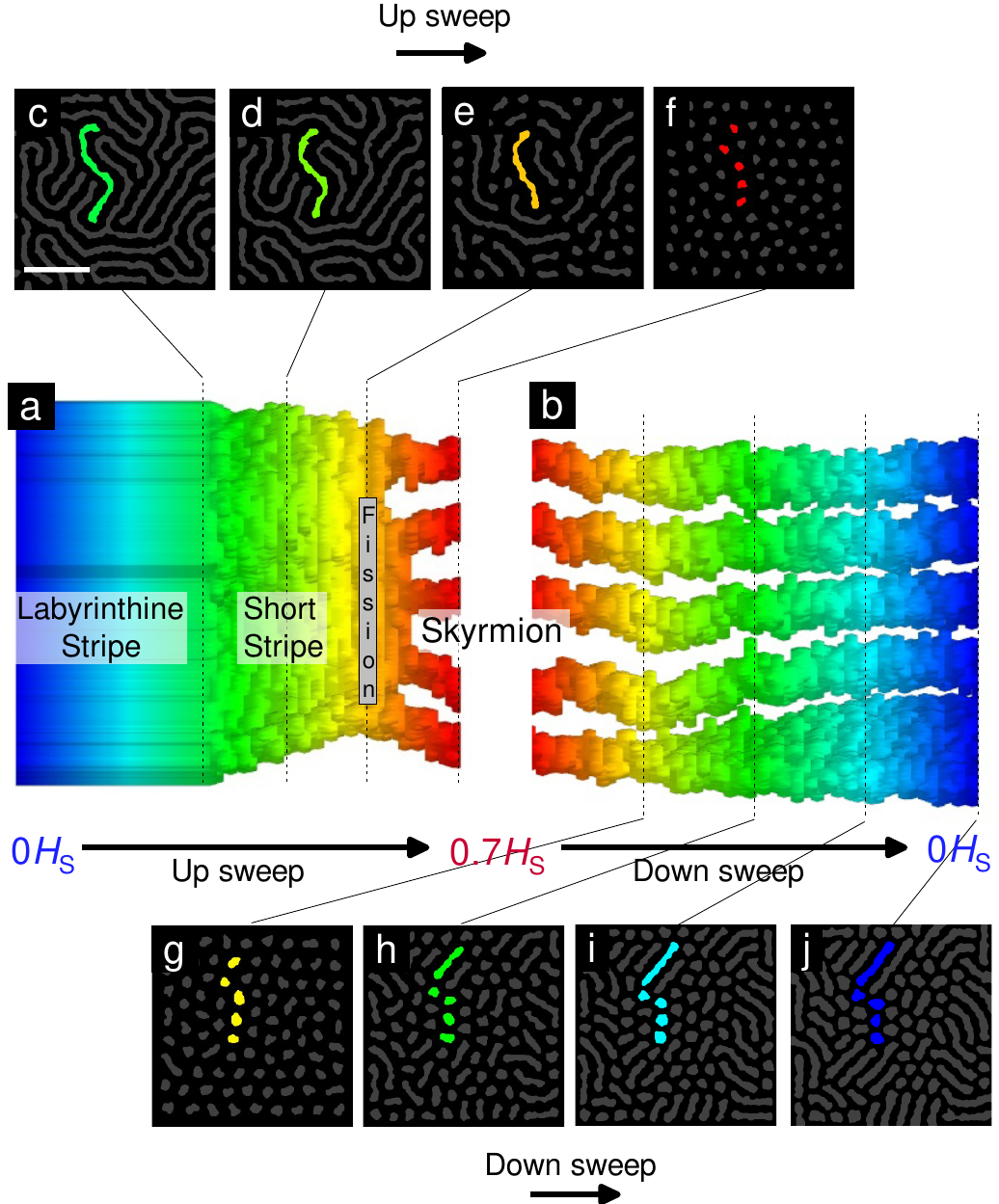}
\par\end{centering}
\noindent \caption[FORC Evolution of Simulated Texture]{\textbf{FORC Evolution of Prototypical Simulated Magnetic Stripe. (a-b)} Stacked 1-dimensional plots, reproduced from manuscript Fig. 4c-d, showing the field evolution of a prototypical stripe along the FORC up-sweep (a) and down-sweep (b). \textbf{(c-j)} Representative magnetization images used to construct the prototypical evolution along the up-sweep (c-f) and down-sweep (g-j). The texture(s) of interest are highlighted using a color scale (consistent with a-b) corresponding to the field magnitude. \label{fig:Sims_TextureEvol}}
\end{figure}

\paragraph{Field Evolution of Simulated Textures}
Here we describe the procedure used to extract the prototypical FORC evolution of simulated magnetic stripes (manuscript Fig. 4c-d, reproduced in \ref{fig:Sims_TextureEvol}a-b) by analyzing the simulated magnetization images produced from the FORC loop (shown for $H_{{\rm r}}=0.7\,H_{{\rm S}}$). Representative magnetization images are shown for the up-sweep (\ref{fig:Sims_TextureEvol}c-f) and down-sweep (\ref{fig:Sims_TextureEvol}g-j) respectively. We begin on the up-sweep by identifying a magnetic stripe after it has broken off from the labyrinthine state (\ref{fig:Sims_TextureEvol}c, colored texture). On the up-sweep, this stripe reduces in length, and fissions into multiple skyrmions (\ref{fig:Sims_TextureEvol}d-f). On the down-sweep, these skyrmions either grow in size, or elongate into individual stripes (\ref{fig:Sims_TextureEvol}g-j). It is clear from a visual inspection of \ref{fig:Sims_TextureEvol}c-j that the particular domain chosen for tracking is not special and represents the typical field evolution of simulated magnetic domains.

To quantify the evolution, the spatial region comprising the stripe is identified and tracked through all images comprising the field sweep (colored in \ref{fig:Sims_TextureEvol}c-j). It is extracted and collapsed in the direction transverse to the stripe to yield a 1-dimensional data set. These 1D data are then stacked horizontally, and the resulting field evolution of their morphology (\ref{fig:Sims_TextureEvol}a-b) shows the irreversibility of the stripe $\rightarrow$ skyrmion transition.

\begin{figure}[h]
\begin{centering}
\includegraphics[width=4.2in]{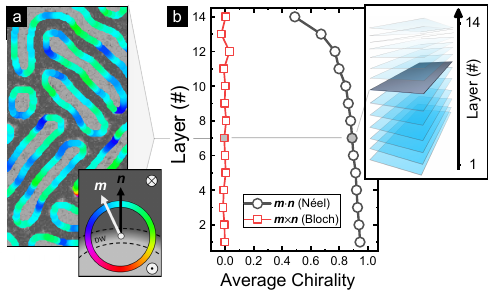}
\par\end{centering}
\noindent \caption[Layer dependence of simulated domain wall (DW) chirality]{\textbf{Layer dependence of simulated domain wall (DW) chirality. (a)} Representative cropped region from the middle layer (7$^{th}$) of the simulated 14-layer stack following zero field micromagnetic simulations performed using Fe(4)/Co(6) parameters. Grayscale colormap represents OP magnetization, and colored ribbons show the angle of IP magnetization ($\textbf{m}$) of DWs relative to the DW normal vector ($\textbf{n}$), as defined in the cartoon inset. \textbf{(b)} Spatially averaged degree of N\'{e}el (defined as $\textbf{m}\cdot \textbf{n}$) and Bloch (defined as $\textbf{m}\times \textbf{n}$) chiralities of DW magnetization across the field-of-view of (a) as a function of layer number, \textit{N}. Inset shows a schematic of the simulated stack.\label{fig:Layer_Chirality}}
\end{figure}

\paragraph{Layer Dependent Chirality}
Previous works on multilayers with low or intermediate DMI ($D \lesssim$ 1 mJ/m$^2$) have reported magnetic textures with layer-dependent, or ``hybrid'' chirality of domain walls (DWs) \citep{Legrand2018,Dovzhenko2018}. The increased prominence of out-of-plane (OP) magnetostatic interactions requires OP flux closure of DWs. This manifests as layer-dependent variation of DW helicity between Bloch and N\'{e}el types. In \ref{fig:Layer_Chirality} we examine the layer-wise DW chirality variation for a simulated zero field configuration with Fe(4)/Co(6) parameters.  The layer dependent N\'{e}el (Bloch) character is given by the scalar (cross-) product of the DW magnetization ($\textbf{m}$) with the DW normal vector ($\textbf{n}$). \ref{fig:Layer_Chirality}(b) indicate that -- notwithstanding some variation across 14 stack repeats -- the average helicity of our spin textures would remain as N\'{e}el-type. Therefore, the broad agreement between simulated and experimental irreversibility (manuscript Fig. 2 and 3) suggests that layer-dependent effects may not play a prominent role in our work. 

\noindent \clearpage{}

\noindent

\noindent \bibliographystyle{apsrev4-2}
\bibliography{SkFORC}

\end{document}